\documentclass[aps,preprint,preprintnumbers,nofootinbib,showpacs]{revtex4}
\usepackage{graphicx,color,amsmath}
\begin{document}
\title{Neutral Pion-like Resonances at Photon Colliders}
\author{Kingman Cheung and Hsin-Wu Tseng}
\affiliation{
Department of Physics and NCTS, National Tsing Hua
University, Hsinchu, Taiwan, R.O.C.
}
\date{\today}

\begin{abstract}
Two photons can annihilate into a neutral pion-like resonance via the 
anomaly coupling, just like $\pi^0\gamma\gamma$ in QCD.  In some
strongly interacting electroweak symmetry breaking models, e.g.,
technicolor type models, there often exist neutral pion-like resonances.
TeV photon colliders have a strong capability to discover such particles,
because the standard model background in photon scattering goes 
through box diagrams and is therefore highly suppressed.  
In this study, 
we perform a signal-background comparison.  We show that $e^+ e^-$ 
linear colliders running in $\gamma\gamma$ mode can discover such 
neutral-pion-like resonances with a decent sensitivity.
\end{abstract}
\preprint{}
\maketitle

\section{Introduction}
The major goal of the next generation collider experiments is to explore
the mechanism of electroweak symmetry breaking (EWSB).  In the standard
model (SM), a Higgs doublet field is responsible for EWSB and gives rise 
to an elementary Higgs boson.  However, the gauge hierarchy problem 
arises from the fact that the radiative correction to the Higgs boson
mass has a quadratic divergence, which demands a finely tuned cancellation
of order of $10^{-16}$ between the bare mass term and the correction terms.
Such a fine tuning problem has motivated a lot of new physics. A particular
class of solutions involves some strong dynamics at the TeV scale.  The 
Higgs boson could then be replaced by a condensate due to the new dynamics.
Technicolor \cite{TC,walk}, topcolor \cite{topcolor}, and topcolor assisted 
technicolor \cite{ttc} models are typical examples of this kind.  
The recent Higgsless models also become strongly coupled at a few TeV 
 \cite{nomura} where some strong dynamics appears.

One of the new features of these strongly interacting models is the 
existence of new mesons and baryons bounded by the new interactions.
In technicolor models often  singlets and doublets of
techni-fermions are proposed.  
These techni-fermions can form bound states via the technicolor
interactions.  If the normal QCD theory is borrowed and rescaled to the
electroweak scale, we would have many techni-mesons and techni-baryons.
As in QCD the neutral pion $\pi^0$ is very unique because of the anomaly
coupling, through which the $\pi^0$ decays into a pair of photons almost
100\%.  Therefore, in technicolor models there are also neutral techni-pions
that couples to a pair of gauge bosons via the anomaly couplings, 
in particular the coupling to a pair of photons.  
In collider experiments,
the neutral techni-pion once produced will decay into a pair of photons
with a sharp invariant-mass peak, which is a very unique
signature for neutral techni-pions.  
Here we consider two photon collisions at TeV scale, which is very unique 
in probing for some neutral pion-like resonances of some new physics 
models.
Some previous studies of technimesons at photon collisions can be found
in Refs. \cite{oldref}.   

\begin{figure}[t!]
\centering
\includegraphics[width=4in]{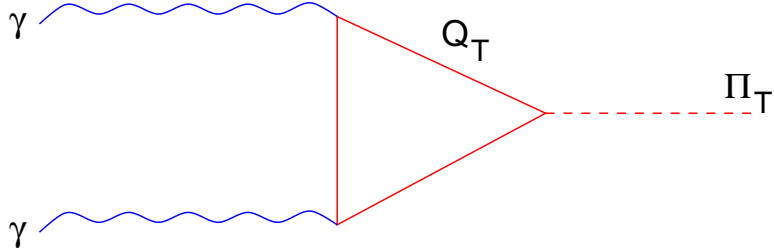}
\caption{\small \label{anomaly}
Feynman diagram for the anomaly-type coupling of the neutral techni-pion.}
\end{figure}

In one simple technicolor model,
the Technicolor Straw Man model \cite{straw}, 
the lightest techni-mesons are constructed solely from the lightest
techni-fermion doublet $(T_U, T_D)$, from which isotriplets
$\rho_T^{0,\pm}, \pi_T^{0,\pm}$ and isosinglets $\pi^{'0}_T, \, \omega^0_T$
can be formed.  In particular, the neutral $\pi^0_T$ and $\pi^{'0}_T$
have an anomaly-type coupling to a pair of photons as 
shown in Fig.~\ref{anomaly}.

In this paper, we specifically work on two models: (i) a rescaled QCD model
\cite{rescale} and (ii) the low-scale technicolor model \cite{low}.  
In the rescaled QCD model, the 
anomaly coupling of the techni-pion $\pi^0_T$ is the rescaled version of the
usual QCD, i.e., the $\pi^0$ decay constant is rescaled by the 
factor $v/f_{\pi^0}$, where $v=246$ GeV and $f_{\pi^0}=130$ MeV \cite{pdg}.
In this particular rescaled QCD model, the $\pi^0_T$
couples to $\gamma\gamma$, $\gamma Z$, and $ZZ$ through the anomaly.  
\footnote{
If the techni-fermions that run in the triangular loop is a 
color singlet, then the techni-pion will not couple to $gg$.
}
We will give more details and formulas in the next section.

In the low-scale technicolor model that we consider is a multi-scale 
technicolor model \cite{low}.  Quark and lepton masses are generated 
by broken extended technicolor gauge interactions in the walking technicolor
model\cite{walk}.  The walking technicolor coupling runs very slowly up to the
extended technicolor gauge scale (a few hundred TeV) 
by including a large number of techni-fermions.  Such a slowly running coupling
allows quark and lepton masses as large as a few GeV to be generated by the
extended technicolor gauge interactions at a few hundred TeV scale.  
There are two types of techni-fermions that condense at widely separated 
scales.  The upper scale is set by $v=246$ GeV while the low scale is
roughly given by $f_{\pi_T}= v/\sqrt{N_D}$, where $N_D$ is the number of 
techni-fermion doublets.  The techni-hadrons associated with the low scale
are of immediate interests at colliders. 
Note that the extended technicolor gauge interaction only generates a few GeV 
to the top quark.  The top mass is, however, generated by another so
called topcolor interaction \cite{topcolor}.
The techni-pions will couple to normal quarks and leptons through the
extended technicolor gauge interactions.  The couplings are Higgs-like
however, and
so the neutral techni-pion will decay into the heaviest possible fermion
pair, e.g.,
\[
 \pi_T^0 \to b \bar b  \;\;\; {\rm or} \;\;\; t \bar t \;,
\] 
depending on the mass of the techni-pion.  Thus, the decay of 
the neutral techni-pion in this model also includes
$gg$, $b\bar b$, $c\bar c$, etc.

The organization is as follows.  In the next section, we describe
the production of and in Sec. III the decay of 
the neutral techni-pion in the rescaled QCD model and the low-scale 
technicolor model in photon collisions.  We conclude in Sec. IV.

\section{Production}

The general form of the anomaly coupling of the neutral techni-pion 
$\pi^0_T$ to two gauge bosons $G_1, G_2$ is given by
\begin{equation}
\label{1}
{\cal M} = N_{TC} {\cal A}_{G_1 G_2} \frac{g_1 g_2}{2\pi^2 f_{\pi_T}}\,
\epsilon_{\nu}\epsilon_{\lambda}
\epsilon^{\nu\lambda\alpha\beta}P_{1\alpha}P_{2\beta} \;,
\end{equation}
where $\epsilon_\nu(P_1)$ and $\epsilon_\lambda(P_2)$ are the polarization
4-vector of the gauge bosons $G_1$ and $G_2$, respectively.  Here 
${\cal A}_{G_1 G_2}$ is the anomaly factor, 
$N_{TC}$ is the number of technicolors, 
$g_i$'s are the gauge couplings of the gauge
bosons, and $f_{\pi_T}=\frac{v}{\sqrt{N_D}}$ is the decay constant of the
techni-pion.
The differential cross section for
$\gamma\gamma \to \pi^0_T$ is given by
\begin{equation}
d \hat{\sigma}= \frac{1}{2 \hat s} \overline{\sum}\left |{\cal M} \right|^{2}
   (2\pi)^{4}\delta^{(4)}(P_1+P_2-P_{\pi_T}) \;
 \frac{d^{3}\vec{P}_{\pi_T} } {(2\pi)^{3} 2 P^0_{\pi_T}}  \;,
\end{equation}
where $P_1$, $P_2$, and $P_{\pi_T}$ are the 4-momenta of the incoming 
photons and the outgoing $\pi^0_T$, respectively,
$\overline{\sum} |{\cal M}|^2$ is the spin-averaged amplitude squared, and
$\hat{s} = (P_1 + P_2)^2$.
The scattering amplitude for $\gamma\gamma \to \pi^0_T$ is obtained from
Eq. (\ref{1}) by specifying the gauge group:
\begin{equation}
i{\cal M}(\gamma\gamma\to \pi_T^0) =i
N_{TC} {\cal A}_{\gamma\gamma}\frac{ e^2}{2\pi^{2}f_{\pi_T}}
\epsilon_{\nu}(P_1) \epsilon_{\lambda}(P_2) 
\epsilon^{\nu\lambda\alpha\beta}P_{1\alpha}P_{2\beta} \;.
\end{equation}
After squaring and averaging over initial polarizations, we obtain
\begin{equation}
\overline{\sum}
\left| {\cal M}\right|^{2}
=\frac{1}{2} \left(N_{TC} {\cal A}_{\gamma\gamma}
\frac{ e^2}{\pi^{2}f_{\pi_{T}} }
\right)^2  (\hat s/4)^2 \;.
\end{equation}
We can then integrate to obtain the cross section as
\begin{equation}
\hat{\sigma}(\gamma\gamma\to\pi_{T}^{0})=\frac{\pi
m_{\pi_T}}{64} \left (N_{TC} {\cal A}_{\gamma\gamma}\frac{ e^2 }{\pi^{2}
f_{\pi_{T}}} \right)^{2} \delta^{(0)}( \sqrt{\hat s}-m_{\pi_{T}}) \;.
\end{equation}
To obtain the realistic cross section $\sigma(\gamma\gamma\to\pi_{T}^{0})$
at an $e^+ e^-$ collider,
we convolute the subprocess 
cross section $\hat{\sigma}(\gamma\gamma\to\pi_{T}^{0})$ 
with the photon luminosity function,
\begin{equation}
\sigma(\gamma\gamma\to\pi_{T}^{0})=\int_{x_{1min}}^{x_{max}}
\!\int_{x_{2min}}^{x_{max}} F_{\gamma/e}(x_1)
F_{\gamma/e}(x_2) \hat{\sigma}(\gamma\gamma\to\pi_{T}^{0}; \;
\hat{s}=x_1 x_2 s) dx_1 dx_2 \;.
\end{equation}
In this paper, $\sqrt{s}$ always refers to the center-of-mass
energy of the parent $e^{+}e^{-}$ collider and $\sqrt{\hat{s}}$ always
refers to the total energy of the two incoming photons. 
The laser backscattering \cite{laser}
is the standard technique to efficiently convert 
an electron beam into a photon beam.  The resulting photon luminosity 
function
$F_{\gamma/e}(x_i)$ is given by \cite{km}
\begin{equation}
F_{\gamma/e}(x_i)=\frac1{D(\xi)} \left[1-x_{i}+\frac1{1-x_i}-\frac{4
x_i}{\xi (1-x_i)}+\frac{4 x_{i}^{2}}{\xi^{2} (1-x_{i})^2} \right] \;,
\end{equation}
where $D(\xi)=(1-\frac4{\xi}-\frac{8}{\xi^2})
\ln(1+\xi)+\frac12+\frac8\xi-\frac1{2(1+\xi)^2}$ and
$\xi\simeq4.8$ in this case for maximal energy conversion. 
Finally, the cross section is
\begin{equation}
\label{8}
\sigma(\gamma\gamma\to\pi_{T}^{0})=
\frac{m_{\pi_T}^2}{2^5 s \pi^3} (\frac{N_{TC} {\cal A}_{\gamma\gamma}
e^2}{f_{\pi_T}})^2 
\;
\int_{x_{min}}^{x_{max}}\frac1{x} F_{\gamma/e}(x) F_{\gamma/e}
 (\frac{m_{\pi_T}^2}{sx}) dx \;.
\end{equation}

In the following, we specifically work on the two models that we described.
The difference between the rescaled model and the low-scale model lies in
the anomaly factor ${\cal A}_{G_1 G_2}$:
\begin{equation}
{\cal A}_{G_1 G_2}=Tr[T^{a}(\{T_1,T_2\}_L+\{T_1,T_2\}_R)] \;,
\end{equation}
where $T_i$ is the generator associated with the gauge boson
$G_i$, and $T^{a}$ is the generator of the axial current associated
with the techni-pion
\begin{equation}
j^{\mu 5 a}=\bar{\psi} \gamma^{\mu} \gamma^{5} T^{a} \psi \;.
\end{equation}
The values of ${\cal A}_{G_1 G_2}$ for the 
rescaled model are essentially the same as the usual QCD 
while the low-scale model involves different values of charges.
Specifically for ${\cal A}_{\gamma\gamma}$, we have
\begin{equation}
{\cal A}_{\gamma\gamma} = Tr(T^a Q^2) = \left \{ \begin{array}{ll}
                \frac{1}{6} & \mbox{ for the rescaled model} \\
                \frac{5}{6} & \mbox{ for the low-scale model} 
                            \end{array} \right . \;,
\end{equation}
which is the consequence of the assignment of charges $Q$.
The electric charge $Q$ of the techni-fermions $T_U$ and $T_D$ is
\[
Q=\left( \begin{array}{cc} Q_{u}&0 \\
0&Q_{d}
\end{array} \right) \;,
\]
where $Q_u$ and $Q_d$ have different values for the two models:

\begin{center}
\begin{tabular}{|c|c|c|}
\hline
          & $Q_u $ & $Q_d$ \\
\hline
rescaled   & $2/3$  & $-1/3$ \\
\hline
low-scale   & $4/3$  & $1/3$ \\
\hline
\end{tabular}
\end{center}

\section{Decay}

Next we have to consider the final states into which the techni-pion decays.
Through the anomaly couplings the neutral techni-pion can decay into
$\gamma\gamma, \, \gamma Z, \, ZZ$ ($gg$ mode is absent if the internal
techni-fermions do not carry color.)  
We particularly choose the $\gamma\gamma$ final
state because of the larger branching ratio and the fact that 
photon does not involve further decay in the detection.
The cross section of $\gamma\gamma \to \pi^0_T \to \gamma\gamma$ is given by,
in the on-shell approximation,
\begin{equation}
\sigma(\gamma\gamma\to\pi_{T}^{0}\to \gamma\gamma)=
\sigma(\gamma\gamma\to\pi_{T}^{0}) \Gamma(\pi_{T}^{0}\to
\gamma\gamma)
\end{equation}
which is valid because the width $\Gamma(\pi^0_T \to \gamma\gamma)$ is
very narrow.
This is a very interesting process because the signal is a tree-level process
while the SM background has to go through box diagrams \cite{box}, which are
naturally suppressed.

Let us first evaluate the branching ratio of $\pi^0_T \to \gamma\gamma$.  In
the rescaled model, the total width is the sum of
$\Gamma(\pi^0_T \to \gamma\gamma)$, $\Gamma(\pi^0_T \to \gamma Z)$,
and $\Gamma(\pi^0_T \to ZZ)$:
\begin{eqnarray}
\Gamma_{total}&=&\frac{c^{2}m_{\pi_T}^{3}}{2^{6} \pi}
+ \frac{c_{1}^{2}(m_{\pi_T}^{2}-m_{Z}^{2})^{3}}{2^{4} \pi
m_{\pi_T}(m_{\pi_T}^{2}+m_{Z}^{2})} 
+\frac{c_{2}^{2}(m_{\pi_T}^{2}-4 m_{Z}^{2})^{2}}{2^{4} \pi
m_{\pi_T}} \nonumber \\
&=&\frac1{2^{4} \pi m_{\pi_T}} \left[\frac{c^{2} m_{\pi_T}^{4}}{2^2}
+\frac{c_{1}^{2} (m_{\pi_T}^{2}-m_{Z}^{2})^{3}}{(m_{\pi_T}^{2}+m_{Z}^{2})} 
+ c_{2}^{2}(m_{\pi_T}^{2}-4 m_{Z}^{2})^{2} \right ]
\end{eqnarray}
where $c=N_{TC} {\cal A}_{\gamma\gamma} \frac{e^2}{2\pi^{2}
f_{\pi_T}}$, $ c_1=N_{TC} {\cal A}_{\gamma Z} \frac{e  g_Z}{2\pi^{2}
f_{\pi_T}}$, $c_2=N_{TC} {\cal A}_{ZZ} \frac{ g^2_Z}{2\pi^{2}
f_{\pi_T}}$, and 
\begin{eqnarray}
{\cal A}_{\gamma\gamma}&=& Tr(T^{a} Q^{2})  \nonumber \\
{\cal A}_{\gamma Z} &=& Tr[T^{a}(T_{3L}+T_{3R}- 2Q \sin^{2}\theta_{w})Q] 
  \nonumber \\
{\cal A}_{Z Z}    &=& Tr[T^{a}((T_{3L}-Q \sin^{2}\theta_{w})^{2}
+(T_{3R}-Q \sin^{2}\theta_{w})^{2})] \nonumber \;.
\end{eqnarray}
The assignment of $Q$ for the two models has been shown in the above.

\begin{figure}[t!]
\centering
\includegraphics[width=5in]{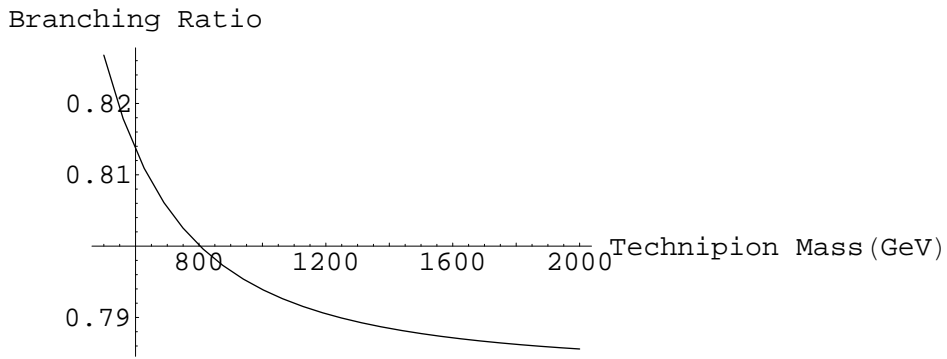}
\includegraphics[width=5in]{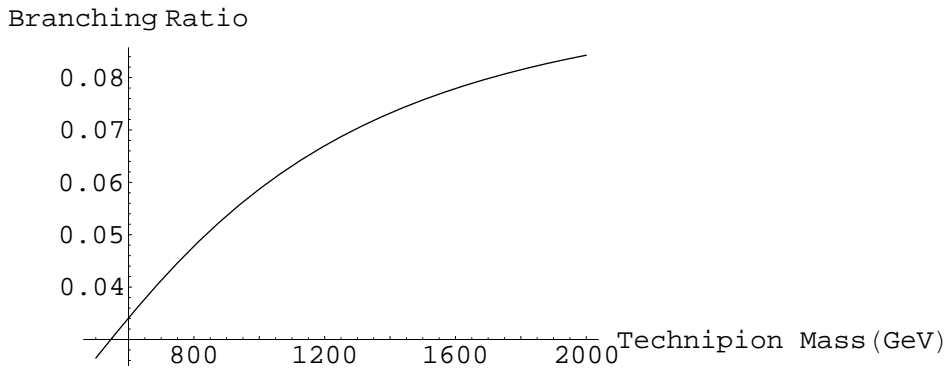}
\caption{\small \label{br}
Branching ratio $B(\pi^0_T \to \gamma\gamma)$ of the techni-pion versus the
techni-pion mass for (a) the rescaled model and (b) the low-scale model.}
\end{figure}

In the low-scale model, we must consider another important mode,
$\pi^0_T \to b\bar b$, the decay width of which is given by
\cite{straw}
\begin{equation}
\left. \Gamma(\pi_{T}^{0}\to b\bar{b}) \right |_{\rm low-scale}
  =\frac1{16\pi f_{\pi_{T}}^{2}}  N_b
P_b C^2_{1b} (m_b+m_b)^{2}
\end{equation} 
where $N_b=3$, $P_b$ is the momentum of the $b$ quark, and $C_{1b}$ 
is a model dependent parameter of order $O(1)$ but the topcolor-assisted
technicolor suggested that $C_{1t} \alt m_b/m_t$, which means
the $t\bar t$ mode is suppressed.
We show the branching ratio $B(\pi^0_T \to \gamma\gamma)$ versus
the techni-pion mass in Fig. \ref{br}.  The rescaled and low-scale models are
shown.  The branching ratio $B(\pi^0_T \to \gamma\gamma)$ for the 
rescaled model decreases with
$m_{\pi_T}$ because the $\gamma Z$ and $ZZ$ modes become less suppressed
in the phase space.  It eventually approaches a stable value when the
techni-pion mass becomes very large. 
On the other hand, $B(\pi^0_T \to \gamma\gamma)$ for
the low-scale model increases with $m_{\pi_T}$ because the $b\bar b$ width
roughly scales as $m_{\pi_T}$ while the $\gamma\gamma$ width scales 
as $m^3_{\pi_T}$.

\begin{figure}[ht]
\centering
\includegraphics[width=3.2in]{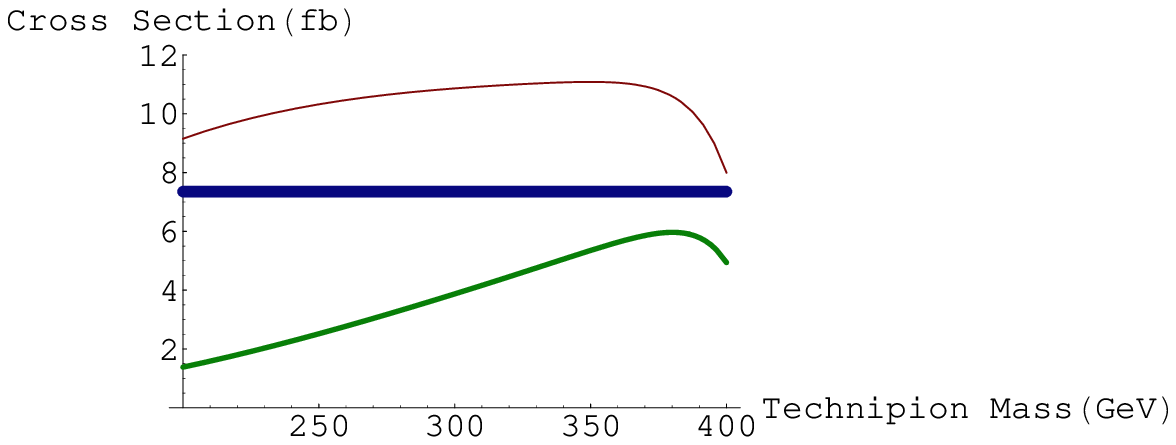}
\includegraphics[width=3.2in]{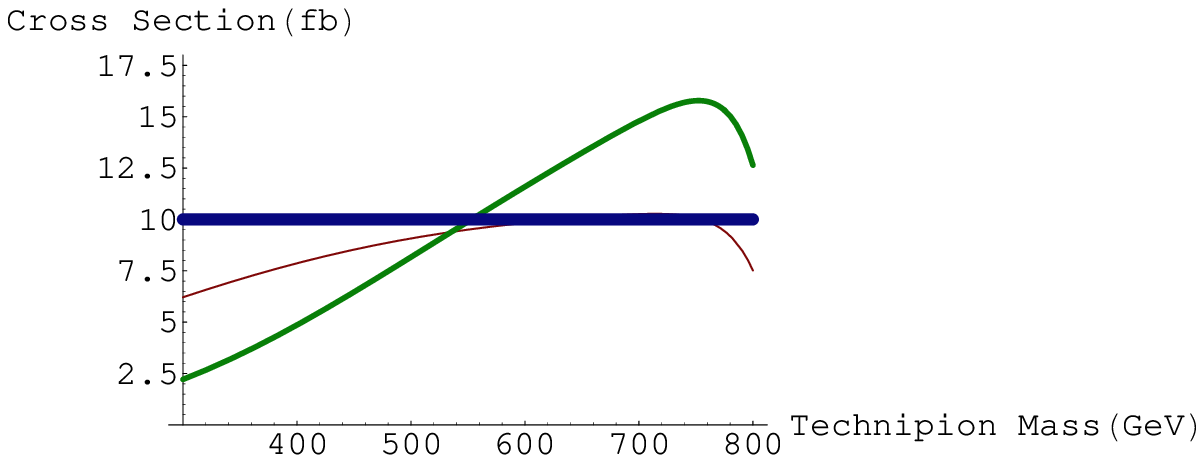}
\includegraphics[width=3.2in]{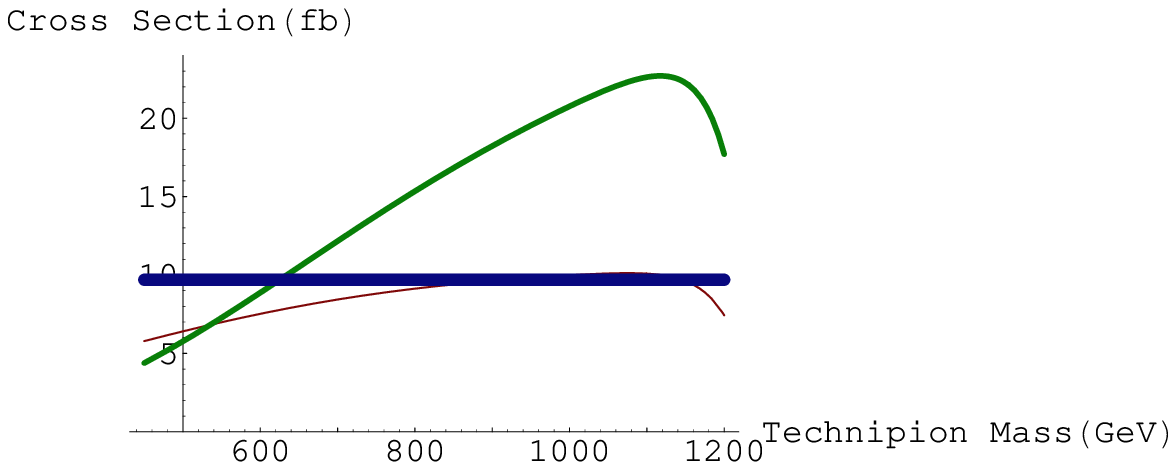}
\includegraphics[width=3.2in]{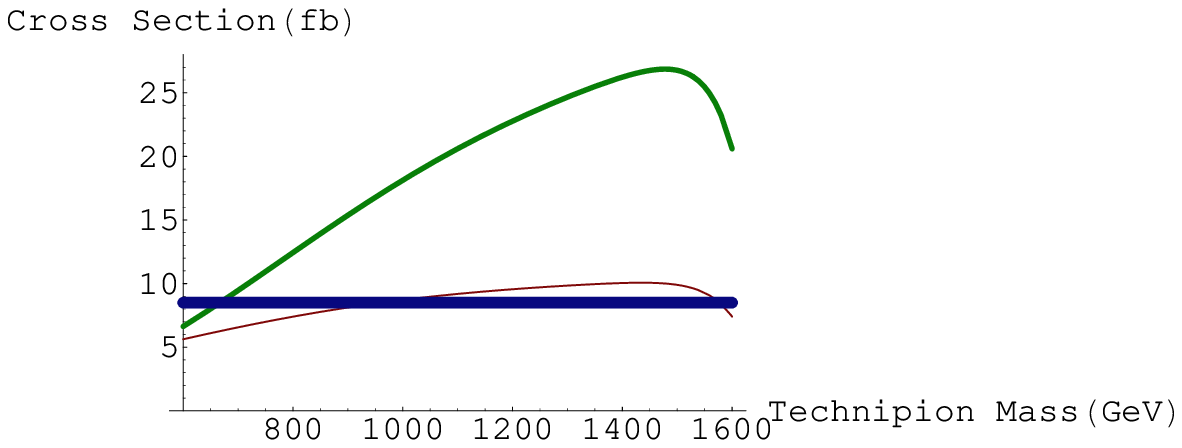}
\caption{ \small \label{fig3}
Production cross sections for the techni-pions of the rescaled model 
(the thin slant line, brown in color) and of the low-scale model 
(the thick slant line, the green line) 
and the SM background (the horizontal blue line)
versus the techni-pion mass at $e^+e^-$ linear colliders running in 
laser backscattering mode.   
(a) $\sqrt{s}_{ee} = 500$ GeV, 
(b) $\sqrt{s}_{ee} = 1000$ GeV, 
(c) $\sqrt{s}_{ee} = 1500$ GeV, and  
(d) $\sqrt{s}_{ee} = 2000$ GeV.
We have used $N_{TC}=4$, $N_{D}=3$, and imposed the kinematic cuts as 
$\frac{M_{\gamma\gamma}}{\sqrt{s}_{ee}}>0.3$ and $|\cos\theta_\gamma| < 
\cos 30^\circ$.
}
\end{figure}

We are now ready to compare the signal cross sections with the SM background.
They are shown in Fig. \ref{fig3} for both the rescaled and low-scale models,
as well as the SM $\gamma\gamma \to \gamma\gamma$ background \cite{box}.  
Some specific choices for $N_{TC}$
and $N_D$ are made.  The cross section $\sigma$ scales with $N_{TC}^2$
and with $N_D$.  We have also imposed kinematical cuts:
\[
\frac{M_{\gamma\gamma}}{\sqrt{s_{ee}}} > 0.3,\;\;\; 
|\cos \theta_\gamma | < \cos 30^\circ
\]
in order to suppress the background, which is very forward and has a 
continuous $M_{\gamma\gamma}$ spectrum.  The SM background is of order of
$O(10)$ fb for $\sqrt{s}_{ee}=0.5-2.0$ TeV.  From Fig. \ref{fig3} the rescaled
model gives a curve which increases very mildly with the techni-pion mass.
This is because the production cross section increases with $m_{\pi_T}$ 
(see Eq. (\ref{8})) but the branching ratio of $\pi^0_T \to \gamma\gamma$ 
decreases with $m_{\pi_T}$ (see Fig. \ref{br}).   On the other hand, 
the low-scale model gives a curve that increases much more rapidly because
the branching ratio of $\pi^0_T \to \gamma\gamma$ also increases with 
$m_{\pi_T}$ (see Fig. \ref{br}.)  Here we have chosen $N_{TC}=4$ and 
$N_D=3$.  Note that $f_{\pi_T} = v/\sqrt{N_D}$.  Both curves eventually dip
down at the upper end because of limitation on the phase space.
It is clear that for a reasonable choice of parameters the techni-pion
signal is rather clean relative to the standard model background. 
In addition, the signal cross section is a few fb to $O(10)$ fb.  The 
event rate with $O(100)$ fb$^{-1}$ luminosity is high enough for a 
feasible search for this kind of signal.  If we are to quantify the 
significance of the signal, we can use $S/\sqrt{B}$, where $S$ and $B$ are the 
number of the signal and background events, respectively.  With an integrated
luminosity of 100 fb$^{-1}$ a signal cross section of 2 fb and a background of
10 fb will give a significance of $6.3$.  Therefore, all the cross sections
shown in Fig. \ref{fig3}  being larger than 2 fb will give sufficiently 
significant signals.   Furthermore, the signal will be a sharp peak determined
by the experimental resolution (the intrinsic width is very small) and 
above the continuum background.

\section{Conclusions}
We have pointed out that the TeV photon-photon collider
is very special in probing for
$O({\rm TeV})$ neutral pion-like resonances, which have a anomaly coupling to 
a pair of photons.  Many extensions of the SM predict the existence of such
resonances.  Famous examples are technicolor models or variants of 
technicolor models.  The advantage of low SM background makes photon colliders
very unique in searching for neutral pion-like resonances.

\section*{Acknowledgments}

This research was supported in part by
the National Science Council of Taiwan R.O.C. under grant no.
NSC 92-2112-M-007-053- and 93-2112-M-007-025-.

\end{document}